\documentstyle[aps,epsf]{revtex}
\begin{document}
\draft
\title{A Semi-Analytical Method For The Evaluation of The Power Spectrum
of a Rotating Observer}
\author{Daniel M\"{u}ller}
\address{Instituto de F\'isica, Universidade de S\~ao Paulo \\
Caixa Postal 66318 -CEP 05389-970 - S\~ao Paulo, S. P., Brasil }
\date{\today}
\maketitle
\begin{abstract}
Abstract: In this letter we propose a semi-analytical method of finding the
power spectrum of a circular moving detector.
\vspace{8 mm}
\end{abstract}
In this letter I consider a {\it monopole} type detector \cite{b1,b2}
interacting with a scalar field $\phi(x)$. The interacting  Lagrangian is,
as usual, $ L_{I}=m(t)\phi(x^{\mu})$. We require both the detector
and the field to obey Schr\"{o}dinger's unitary time evolution. The
physical states are described in the usual way as a particular
representation of the abstract Hilbert space, given by the tensor
product of the detector's and field states. Let us assume that in the
very early past the detector state is some state $|E\rangle$ and that
the field is in the (Minkovswski) vacuum state $|0\rangle$.  As the
detector moves, field and detector will undergo transitions to various
states, in a trajectory dependent fashion. Since one does not care
about the field's final state, the unknown final state of the field
should be traced out from the transition probability for the
transitions  in question. This process yields, in first order of
perturbation  theory the following expression for the transition
probability \cite{b2}
\begin{equation}
P(E) = C(E) \lim_{\tau_{0} \rightarrow \infty} \frac{1}{2
\tau_{0}}\int_{-\tau_{0}}^{+\tau_{0}} \int_{-\tau_{0}}^{+\tau_{0}}
e^{-iE(\tau-\tau\prime)} \langle 0_{M}|\phi(x^{\mu}(\tau))
\phi(x^{\mu}(\tau\prime))|0_{M} \rangle
d\tau d\tau\prime.
\end{equation}
Here $C(E)$ depends on the internal constituency of the detector while
the second term corresponds to the noise this detector is submitted
to, that's to say, on the way the field fluctuates as seen by the
observer in his trajectory $x^{\mu}(\tau)$. As the detector follows a
non-inertial trajectory, there is a  finite probability of being exited.
As   well known since  Unruh's milestone work, for a uniformely
accelerated  trajectory the power spectrum has Planckian form. That is to
say, the detector feels exactly as if  it were immersed in a black body
bath at temperature $T=\frac{\hbar a}{2 \pi c k_{B}}$ where $a$ is  it's
proper acceleration, $c$ the light velocity, $\hbar$ and $k_{B}$ are
Planck's and Boltzman's constant respectively
\cite{b1}.

When the detector follows a  circular trajectory with
uniform angular velocity, then the detector gets also excited, but the
spectrum is no longer Planckian \cite{b3}. Circular motion has proven to be
very
interesting because it can be regarded as an experimental verification of the
Unruh effect trough the depolarization of an electron beam in circular particle
colliders  \cite{b4}. Unfortunatly, this power spectrum was so far obtained via
numerical integration in a case by case fashion. In this letter we propose a
semi-analytical general method and do study some specific examples. It's an
easy exercise to show that for a uniform rotation of radius $R$ and and
velocity
$v$:
\begin{equation}
P(E) = - \frac{a}{8 \pi^2 v \gamma^3} \int_{-\infty}^{+\infty}
\frac{e^{-i\beta Ex}}{x^2 - v^2\sin^2x}dx,
\end{equation}
where $\gamma$ is the usual Lorentz factor,  $\beta=\frac{2v
\gamma}{a}$ and $a=v^2 \gamma^2/R$ is the proper acceleration. The
above integral will be obtained via  the  method of residues. The
transcendental equation
\begin{equation}
z = v \sin z,
\end{equation}
yields the poles labeled $z_{n}$ in the complex plane. The above equation can
be
separated in real and imaginary parts with $z$ represented by $z=x+ i y$,
forming
a system of two equations
\begin{eqnarray}
x&=&v \sin x \cosh y \\
y&=&v \cos x \sinh y.
\end{eqnarray}

When $x=0$ we have two pure imaginary poles that are given by the equation (5),
whose graphical solution we plot in FIG. 1.
\begin{figure}[htpb]
\centerline{
\epsfxsize=10cm
\epsffile{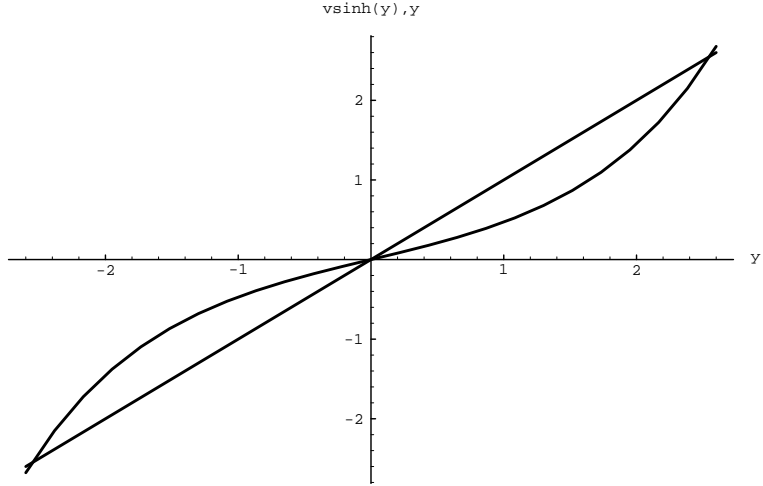}}
\caption{- graphical solution of equation y-vsinhy=0, with v=0.4.}
\end{figure}
For a further inspection of the location of the poles we use polar coordinates
$z= \rho e^{i \theta}$ in (3) neglecting the negative exponential to obtain
(we assume $\rho>>1$)
\begin{eqnarray}
e^{\rho\sin\theta}\cos(\rho \cos\theta)&=&\frac{2\rho}{v}\sin\theta\\
e^{\rho\sin\theta}\sin(\rho \cos\theta)&=&\frac{2 \rho}{v}\cos\theta,
\end{eqnarray}
wich upon squaring and summing one to another gives for $\sin \theta$
\begin{equation}
\sin \theta = \frac{1}{\rho} \log (2/v) + \frac{ \log \rho}{\rho}.
\end{equation}
Equation (8) says that as we move away from the origin
 $\theta \sim n \pi$. Because of the reflection symetry of equation (3) we have
that the poles are near the real axis and symetricaly  located as sketched in
FIG. 2.
\begin{figure}[htpb]
\centerline{
\epsfxsize=10cm
\epsffile{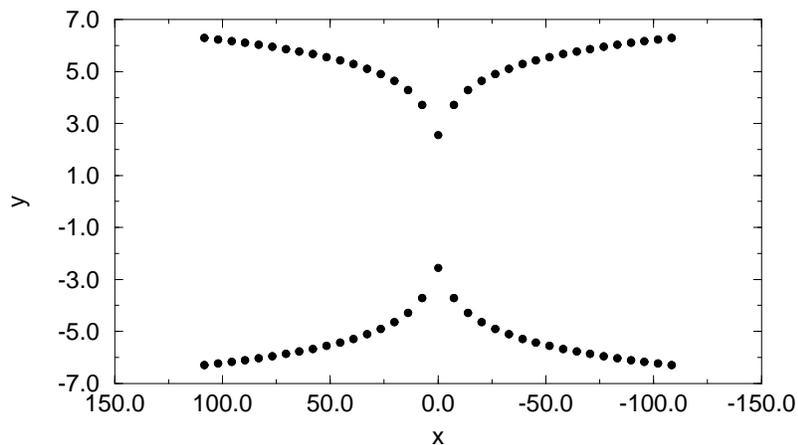}}
\vspace{1cm}
\caption{-location of the poles of vsinz=z for v=0.4 in the complex
plane with the real part represented by x and the imaginary part by y.}
\end{figure}

To obtain the exact location of the other poles, when $x\neq 0$, we have to
solve
\begin{equation}
v \cos x \sqrt{ \frac{x^{2}}{(v \sin x)^{2}} - 1}- arcosh ( \frac{x}{v \sin x}
) = 0.
\end{equation}
wich is obtained in a combination of (6) and (7).

The contour of integration has to be closed from bellow, and the integral is
obtained by summing the first order residues
\begin{equation}
P(E)=- \frac{i v}{8 \pi r \gamma} \sum_{n}^{\infty} \frac{e^{-i E \beta z_{n}}}
{-z_{n}+v^{2} \sin z_{n} \cos z_{n}}.
\end{equation}
 This series turns out to converge rapidly enough. Therefore we have only to
sum over a few poles to obtain a fairly good approximation. The poles are found
in  different regimes:
\begin{itemize}
\item In the relativistic limit $v \rightarrow 1$ the
pure imaginary pole near the origin gives the most important
contribution to the integral \cite{b5}. The denominator of the
integral can be expanded in a power series in z and the analitical
result yields
\begin{equation}
P(E)= \frac{a}{8 \pi \sqrt{3}} e^ \frac{-2\sqrt{3}E}{a},
\end{equation}
\item  Taking  $\sin\theta$ from expression (8) and substituting in right side
of (6)
with the approximation $\cos( \rho \cos \theta )\sim \cos \rho$  since $\theta
\sim
2n \pi$ (we analize only the first quadrant) results
\begin{equation}
\cos\rho=\frac{\log(2/v)}{\rho} + \frac{\log(\rho)}{\rho}.
\end{equation}
For sufieciently large $\rho$ we can expand the above equation in the vicinity
of $\rho_{0}=2n \pi + \pi /2$, to obtain an approximate value for $\rho$,
\begin{eqnarray}
\rho&=& 2n \pi + \pi/2 +\left[ \frac{\log(2/v)}{2n \pi + \pi /2} +
\frac{\log(2 n \pi + \pi/2)}{2n \pi + \pi /2} \right] \nonumber\\
& &\left[ \frac{\log(2/v) + \log(2n \pi + \pi /2) -1}{(2n \pi +
\pi/2)^2} - 1\right]^{-1}.
\end{eqnarray}
\item In the non relativistic limit $v \rightarrow 0$ it follows from
(8) that $y \sim \log(v/2)$. The approximate value for the real part $x \sim 2n
 \pi$
is obtained dividing (4) by (5) and remembering that in this limit $\tanh y
\sim
\pm 1$. Substituition
of these two approximations in $y=\rho \sin\theta$ with $\sin \theta$ given by
(8)
gives $y=\log(v/2) +1/2 \log(4^2n^2 \pi^2+\log(v/2)^2)$. After one iteraction
the result is further improved:
\begin{eqnarray}
&y&=\log(2/v) +1/2 \log \left(4n^2 \pi^2 + \left[
\log(2/v) \right. \right.\nonumber\\
&+&\left. \left.1/2 \log(4n^2\pi^2+(\log (2/v))^2)\right]^2\right).
\end{eqnarray}
\end{itemize}

The approximations used for finding the poles (eqs.(13) and (14)) were
confronted against direct numerical evaluation of the roots of eq. (9)
showing good agreement. Once the poles are obtained they
are substituted in equation (10). In what follows we plot the graphs of $F(E)$,
the spectrum itself, which is obtained from the power spectrum $P(E)$ upon
multiplication by the density of states $E^{2}/4 \pi$.
\begin{figure}[htpb]
\centerline{
\epsfxsize=10cm
\epsffile{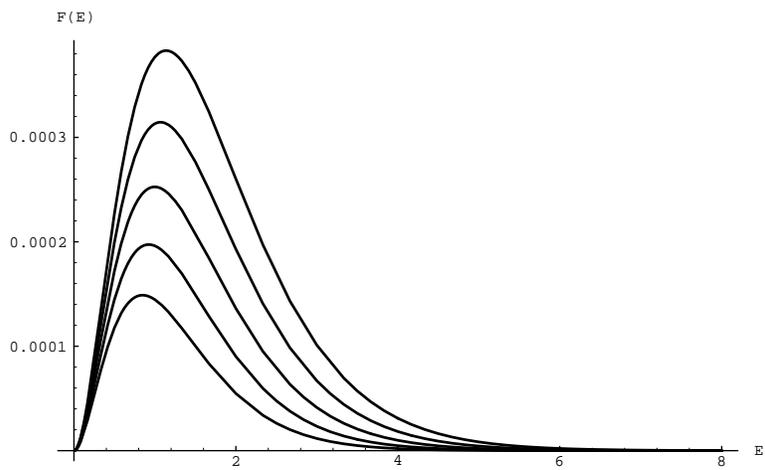}}
\caption{- results obtained with P(E) given by (10) with
v={0.868243, 0.894427, 0.913812, 0.928477, 0.939793, 0.948683}, and
R={2.21057, 2.61532, 3.03826, 3.478, 3.93338, 4.4034}}
\end{figure}
\begin{figure}[htpb]
\centerline{
\epsfxsize=10cm
\epsffile{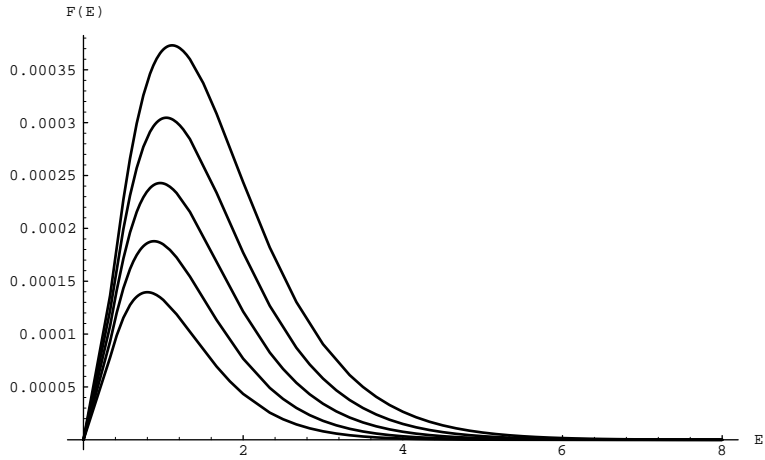}}
\caption{- direct plot of F(E), with P(E) given by equation (11),
with the same values of v and R of FIG.3, showing when the relativistic limit
ceases to be valid.}
\end{figure}

\acknowledgments I'm thankful to M. Schiffer and G. Matsas
for valuable discussions and $CNP_{Q}$ for financial support.


\begin{thebibliography}{99}
\bibitem{b1} Unruh, W. G. (1976), Phys. Rev. {\bf D14}, 870.
\bibitem{b2} DeWitt, B. S. (1979), in General Relativity, ed. S. W. Hawking
and W. Israel (Cambridge University Press), p. 680.
\bibitem{b3} Letaw, J. R. (1981), Phys. Rev. {\bf D23}, 1709.
\bibitem{b4} Bell, J. S. and J. M. Leinaas (1983), Nucl. Phys.
{\bf B212} , 131.
\bibitem{b5} Takagi, S. (1986), Prog. Theo. Phys. Supplement {\bf 88} p. 134.
\end{thebibliography}
\end{document}